\documentclass[prl,twocolumn,superscriptaddress,showpacs,amsmath,amssymb]{revtex4}

\usepackage{graphicx}
\usepackage{latexsym}
\usepackage{amsmath}
\usepackage{amssymb}
\usepackage{amsfonts}
\usepackage{color}
\usepackage{bm}
\usepackage{verbatim}
\usepackage{changes}

\newcommand{\nn}{\mathrm}

\begin{document}
\title{Excess magneto-resistance in multiband superconductors due to the viscous flow of composite vortices}

\author{Artjom Vargunin}
\affiliation{Institute of Physics, University of Tartu, Tartu, EE-50411, Estonia}
\affiliation{Department of Theoretical Physics, The Royal Institute of Technology, Stockholm, SE-10691 Sweden}
\author{Mihail Silaev}
\affiliation{Department of Theoretical Physics, The Royal Institute of Technology, Stockholm, SE-10691 Sweden}
\author{Egor Babaev}
\affiliation{Department of Theoretical Physics, The Royal Institute of Technology, Stockholm, SE-10691 Sweden}

\date{\today}

\begin{abstract} 
By using the time-dependent Ginzburg-Landau theory, we show that extremely diverse experimental data on flux-flow resistivity in multiband superconductors can be qualitatively explained by a composite nature of Abrikosov vortices consisting of elementary fractional vortices in different bands. In composite vortices, the ratio of a core 
 size to electric field relaxation length is found to vary in wide limits depending on system parameters. As a result, the flux-flow 
 magneto-resistance can strongly exceed the single-component Bardeen-Stephen estimation provided that moving vortices generate
  electric field stretching strongly outside the vortex cores.     
\end{abstract}


\maketitle
Recent experimental studies of resistive states in multiband superconductors revealed unusual vortex physics associated with
 the viscous flow of magnetic flux. The magnetic field dependencies of flux-flow resistivity $\rho_\nn{f}$ were found to be qualitatively different from the single-band behaviour established  in classical experiments \cite{Kim} and theoretical works by Bardeen and Stephen \cite{BS}, Tinkham \cite{tinkham}, and Nozieres and Vinen \cite{NV}. 
 
In conventional type-2 superconductors at low temperatures $T\ll T_\nn{c}$ 
the flux-flow resistivity is well described by a linear  Bardeen-Stephen (BS) law $\rho_\nn{f}/\rho_\nn{n}= \gamma \mathcal{B}/H_{\nn{c}2}$, where $\rho_\nn{n}$ is the normal state resistivity, $\mathcal{B}$ is an average magnetic induction, $H_{\nn{c}2}$ is the second critical field and 
$\gamma\approx 1$ is the magneto-resistance coefficient \cite{Kim}. At elevated temperatures $T\to T_\nn{c}$, the 
vortex motion is strongly impeded due to the enhanced electron-phonon relaxation which results in a  
significant suppression of $\rho_\nn{f}/\rho_\nn{n}$ so that magneto-resistance defined by asymptotic at low magnetic fields decreases below the BS value. In the gapless regime 
$\gamma\approx0.69$ \cite{Schmid}. Similarly, the vortex motion becomes more viscous due to the depairing effects resulting from spin-flip scattering at magnetic impurities. In this case $\gamma=0.33$ \cite{Hu}. 
 \begin{figure}[t]
 \includegraphics[trim=20 10 165 625,clip,width=1.0\linewidth]{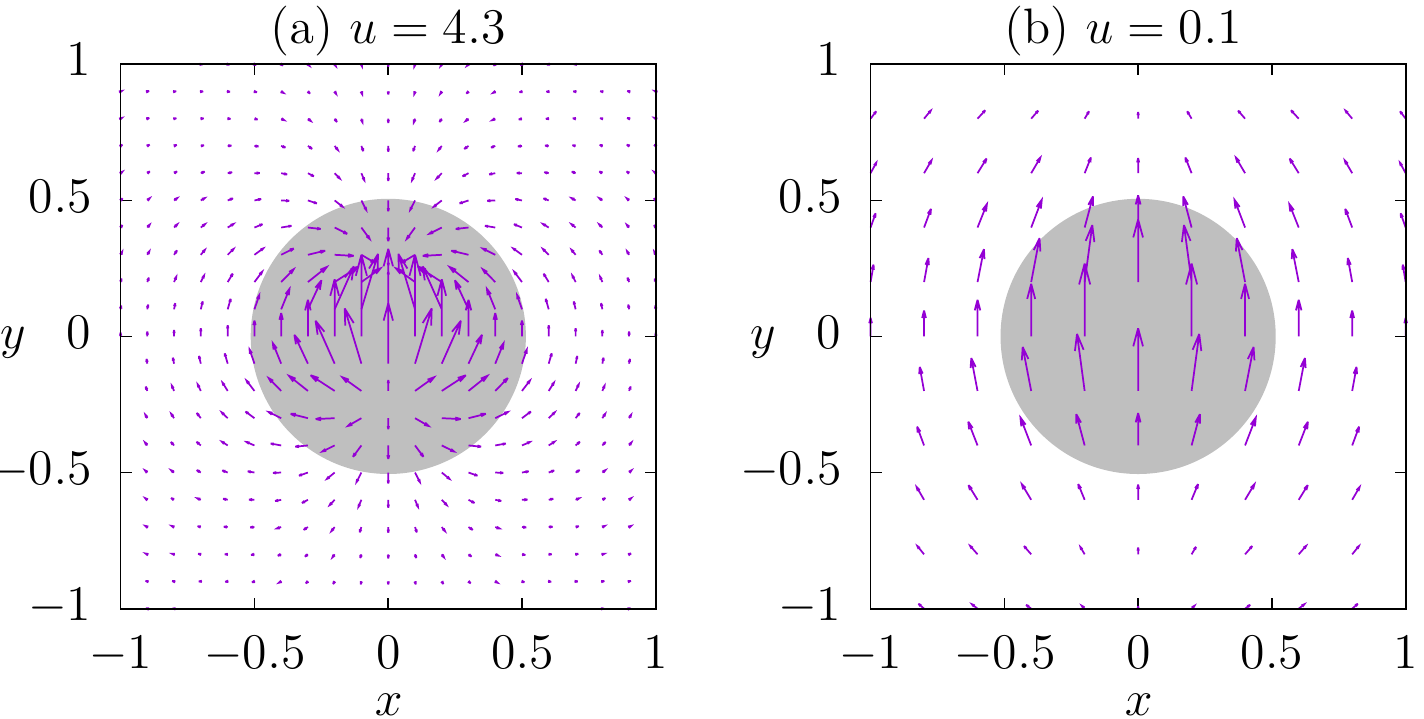}
\caption{Distribution of electric field (vectors) in a two-band superconductor around composite vortex 
 line oriented along the $z$ axis. Vortex core size (gray area) is defined by the boundary where
 the gaps recover
  $0.95 \%$ of the bulk value. The panels differ only in the ratio of diffusion coefficients
 (a) $D_2/D_1=0.5$, $u=4.3$ and (b) $D_2/D_1=70$, $u=0.1$.  The parameter $u$ is defined in text. }\label{f1}
 \end{figure}

 In contrast to the conventional single-band behaviour, multiband superconductors $MgB_2$ \cite{exp1} and $Li111$ \cite{exp3,exp4}
 were found to have the flux-flow magneto-resistance increased   
 above BS-value line $\gamma>1$ so that  $\rho_{f}/\rho_n>\mathcal{B}/H_{c2}$.
 The experimentally found dependencies $\rho_f(\mathcal{B})$ have a steeper growth in the low-field region with 
 an enhanced magneto-resistance $\gamma\approx 1.4$. For $P$-$Ba122$ \cite{exp2} and $P$-$Sr122$ \cite{exp5} systems, even larger slopes were obtained $\gamma\approx 2.5$. 
Moreover, the opposite behaviour of smaller magneto-resistance was also observed in experiments with $FeSeTe$, that reported  
$\gamma\approx 0.7$ \cite{exp6}.
 
  In this paper we show that the enhanced flux-flow magneto-resistance can be explained by a composite nature 
 of vortices in multiband supercondutors. These complex topological excitations consist of several singularities corresponding to phase windings of components of the order
 parameter in different superconducting bands.  
 Such composite objects can be considered as bound states of several fractional vortices \cite{babaev2}. In most cases 
 the equilibrium state corresponds to co-centred fractional vortices although they can split under the action of 
 fluctuations \cite{babaev3,babaev2004phase}, external drive \cite{lin} or due to the interaction with other vortices and sample 
 boundaries \cite{dao,silaev}. 
 
 The viscous motion of composite vortices under the action of an external Lorentz force is determined simultaneously 
 by the non-equilibrium processes in several superconducting bands. The presence of additional conducting bands with 
 smaller gap amplitudes increases the total density of normal electrons trapped within vortex cores. This weakens the 
 screening of electric field, which can stretch out of the vortex core at distances far exceeding the coherence length, 
 cf panels in Fig. \ref{f1}. As a result of such a  non-trivial interplay of the 
 vortex core size and the electric field relaxation length, the value of magneto-resistance $\gamma$ can change in wide
 interval. Such a behaviour is in high contrast to single-band superconductors characterized in the gapless regime by 
 a universal value $\gamma\approx 0.69$ \cite{Schmid}.
  
 The non-equilibrium processes related to the vortex motion are known to be rather diverse. Different dissipation 
 mechanisms can play the dominating role depending on the range of temperatures and magnetic fields. 
 In this paper we consider an $s$-wave superconductor \footnote{We focus on the standard $s$-wave case. Different physics emerges in 
 the $s+is$ state due to time-reversal symmetry breakdown, see 
 M. Silaev, and E. Babaev, Phys. Rev. B {\bf 88}, 220504(R) (2013)} at temperatures in the immediate vicinity of the critical one 
 given by $(T_\nn{c}- T) \ll \hbar\tau_\nn{ph}^{-1}$, where $\tau^{-1}_\nn{ph}$ is the electron-phonon scattering rate. 
 Within such temperature  interval, gapless superconducting states are realized as a result of the inelastic interactions with phonons. 
 
 Non-equilibrium effects in gapless superconductors can be described by the time-dependent Ginzburg-Landau (TDGL) theory. 
 We start from the microscopic weak-coupling model of a dirty two-band superconductor described by the $2\times 2$ matrix of 
 coupling constants $\hat\Lambda = \left( \lambda_{11} \; \lambda_{12} \atop \lambda_{21} \; \lambda_{22}\right)$ and diffusion coefficients $D_k$ in each superconducting band ($k=1,2$). We derive TDGL generalization for a two-band superconductor formulated in terms of the two-component complex field  $(\psi_1,\psi_2)$, see Supplementary Material.  
 
 The two-band Ginzburg-Landau model is an expansion in several small parameters \cite{Tilley}: small gaps and gradients, not to be confused
 with the expansion in single small parameter $\tau=1-T/T_\nn{c}$. For conditions when it holds and accurately
 approximates microscopic theory see \cite{silaev2}. The subdominant component can alter substantially the 
 magnetic properties \cite{babaev0,babaev5,carlstrom2011type,silaev2,babaev4}. However, for ordinary 
 $s$-wave superconductors that undergo a single
 second-order  phase transition, the parameter $\tau$ must become the smallest one
 sufficiently close to $T_\nn{c}$ since the system breaks only $U(1)$ symmetry.
 Then the model is well approximated by a single-component 
 Ginzburg-Landau theory \cite{Kresin}. In this paper we are interested in the regime very close to $T_\nn{c}$. Then the second component can be excluded by
 projecting the fields $\psi_{1,2}$ to the 
  eigenvector corresponding to the maximal eigenvalue   
 of the coupling matrix $\hat \Lambda $, see detailed discussion in \cite{silaev2}. 
 To implement such a projection we use the ansatz
 $(\psi_1,\psi_2)^T= \psi(\mu, 1)^T$, where $\mu=(\lambda_-+\lambda_0)/(2\lambda_{21})$, $\lambda_-=\lambda_{11}-\lambda_{22}$ and $\lambda_0=\sqrt{\lambda_-  + 4\lambda_{12}\lambda_{21}}$.
 Here the common complex field $\psi$ is a Landau order parameter
 corresponding to a $U(1)$ symmetry breaking in the two-band system. 

 Applying the transformation outlined above,
 we obtain the effective single-component TDGL equation for the two-band system
  \begin{align} \label{Eq:AppTDGL}
 &\Gamma \left(\partial_t+ 2ie\varphi/\hbar \right)\psi=-\delta_{\psi^\ast}\cal F, \\
 & {\cal F}= \int d^3{\bm R}\left(\alpha |\psi |^2 + K |(\nabla-2\pi i\bm A/\phi_0)\psi|^2 +\frac{\beta}{2}|\psi|^4\right).\nonumber
 \end{align} 
Here $\varphi$, $\bm A$ are scalar and vector potentials, $\phi_0$ is flux quantum and expansion coefficients are given by
 \begin{align}\label{eff_coef}
 &\alpha=-\nu\tau,\qquad\beta=7\zeta(3)(\nu_2+\nu_1\mu^4)/(8\pi^2 T_\nn{c}^2),\nonumber\\
 &K=\pi\hbar\nu D/(8T_\nn{c}),\qquad\Gamma= \pi\hbar \nu/ (8T_\nn{c}),
 \end{align}
 where $\nu_k$ is density of states in each band, 
 $\nu=\nu_2+\nu_1\mu^2$ and $D=(\nu_2D_2+\nu_1D_1\mu^2 )/\nu$.
In this approximation the order parameter is a composite field that describes joint contribution of two superconducting bands.

  The non-equilibrium response described by Eq. (\ref{Eq:AppTDGL}) 
 is determined by the parameter $u=\xi^2/l_\nn{e}^2$ where $\xi$ and $l_\nn{e}$ are the coherence length
 and electric field relaxation length, respectively. The value of coherence length $\xi=\sqrt{-K/\alpha}$ 
 can be obtained from the equilibrium GL functional. 
The electric field relaxation length $l_\nn{e}$ can be found from the TDGL theory supplemented by 
 the charge conservation law $\nabla\cdot ({\bm j}_\nn{s}+{\bm j}_\nn{n} )=0$, where  
 ${\bm j}_\nn{s} $ is the superconducting current and ${\bm j}_\nn{n}=\sigma_\nn{n} {\bm E}$ is the normal current given by electic field ${\bm E}$ and normal state conductivity
 $\sigma_\nn{n}=2e^2\sum_k\nu_kD_k $. By introducing gauge invariant scalar potential $\Phi=\varphi+\hbar\partial_t\theta/(2e)$, we write Eq. (\ref{Eq:AppTDGL}) in the form $\nabla\cdot {\bm j_n} = -8e^2 \Gamma |\psi|^2 \Phi/\hbar^2$ so that screening of scalar potential is described by equation
$l_\nn{e}^2\nabla^2\varphi= f^2\Phi$, where 
 $l_\nn{e}= \sqrt{\hbar^2 \sigma_n / (8e^2\Gamma \psi_0^2)}$ and $\theta$, $f$ are order-parameter phase and its amplitude scaled by bulk value $\psi_0 = \sqrt{-\beta/\alpha}$.

 In contrast to single-band superconductors which have the fixed value 
 of $u\approx 5.79$ \cite{Ivlev,kopnin}, in multi-band compounds $u$ strongly depends on the system parameters.
 Using the expressions for $\xi$ and $l_\nn{e}$ obtained above together with the GL coeffitients (\ref{eff_coef}) we obtain
 \begin{align}\label{u}
 u=5.79\frac{(\nu_1\mu^2+\nu_2)(\nu_1D_1\mu^2+\nu_2D_2)}{(\nu_1\mu^4+\nu_2)(\nu_1D_1+\nu_2D_2)}.
 \end{align}
 As long as the diffusion coefficients $D_{1,2}$ can differ in orders of magnitude in realistic compounds
 \cite{diffus1,diffus2,diffus3,diffus4}, the latter expression can change in wide limits ranging from $u\ll 1 $ to $u\gg 1$.
   
 Let us consider several characteristic cases. First, we can assume that the interband pairing is weak $|\lambda_{12}|\ll \lambda_{kk}$.
 In this case $\mu\gg1$ so that Eq. (\ref{u}) yields $u=5.79\nu_1D_1/(\nu_1D_1+\nu_2D_2)$. This expression does 
 not exceed the universal single-band numerical value realized when $\nu_2D_2\to 0$. Qualitatively this result 
 means that the electric field is screened less effectively due to 
 presence of additional band with enhanced concentration of normal electrons. 
 In the opposite case of interband-dominated pairing interaction provided that $\mu\sim 1$, the single-band value of $u$ is 
 recovered irrespective  of the ratio of diffusion coefficients.
 
 Resistive states of superconductors containing composite Abrikosov vortices is dramatically affected by an interplay
 between lengths $\xi$ and $l_\nn{e}$. The rate of energy dissipation induced by moving vortices is determined by 
 the ratio of a vortex core size which is proportional to $\xi$ and the size of a domain where the electric field generated 
 around moving vortex is concentrated. In the usual BS approach it is assumed that these scales are identical while the theory 
 of gapless single-band superconductors with $u=5.79$ determines stronger electric field localization. This leads to smaller 
 values of flux-flow magneto-resistance $\gamma=0.69$ as compared to the BS estimation $\gamma\approx 1$.
 
\begin{figure}[t!]
 \includegraphics[trim=20 40 125 610,clip,width=1.0\linewidth]{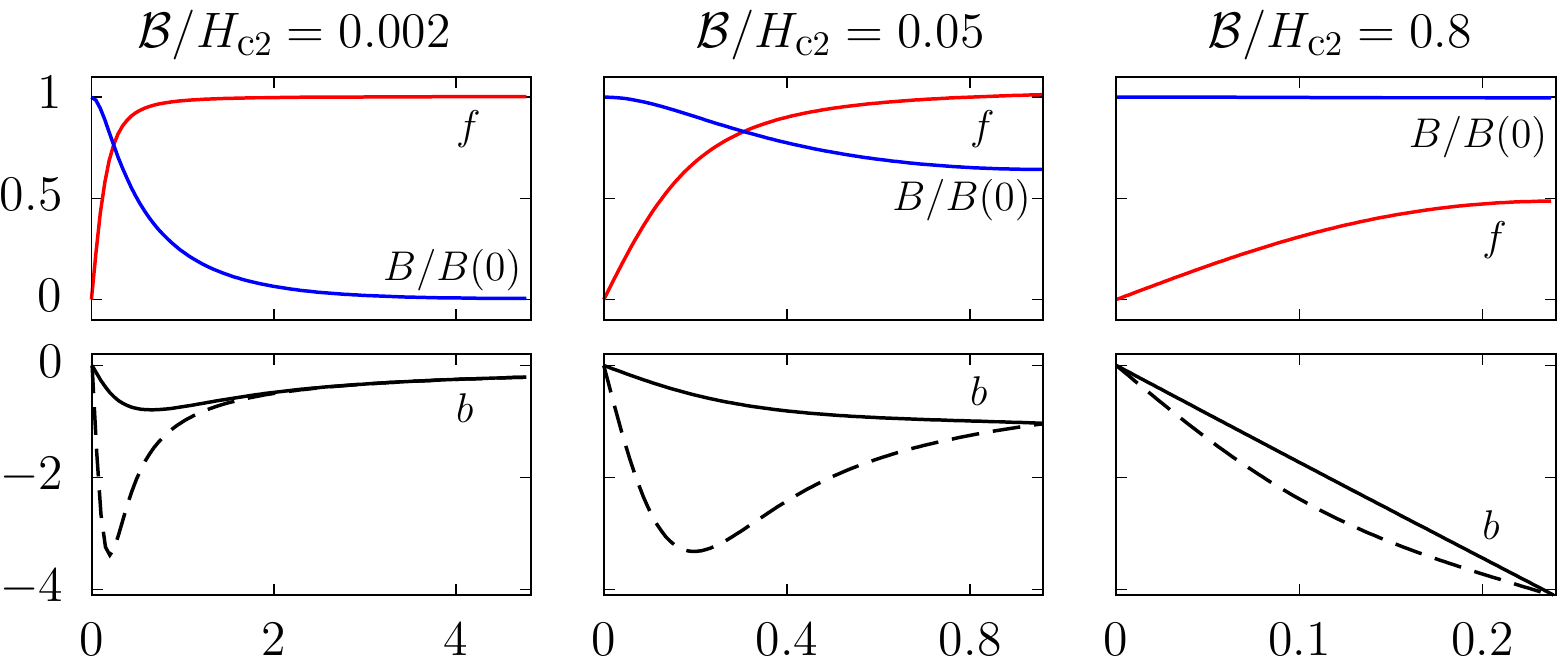} 
 \caption{\label{f2} 
 (Color online) Upper row: The distributions $f(r)$ and $B(r)/B(0)$ for the order parameter and magnetic field 
  is depicted for different vortex concentrations set by an average magnetic induction. Lower row: Distributions of scalar potential $b(r)$. For comparison, cases with $u=0.1$ (black solid) and $u=5.7$ (black dashed) are shown.  }
 \end{figure}

 In multiband systems one can obtain qualitatively different regimes when the scale of electric field localization around moving 
 composite vortices strongly exceeds the vortex core size. The resulting flux-flow resistivity can significantly exceed the 
 BS estimate. We study this effect by calculating vortex structures and electric field created by moving vortices arranged in 
 the regular lattice. To model the vortex array at finite values of magnetic induction we use standard  circular cell approximation. 
 In this approach the hexagonal unit cell of the triangular vortex lattice is replaced by a circle, where the order parameter and 
 magnetic field distributions are assumed to be axially symmetric. The circular cell radius $R_\nn{c}$ is uniquely defined by averaged magnetic induction $\cal B$ due to flux quantization 
 $\pi R_\nn{c}^2=\phi_0/\mathcal{B}$. 
Several vortex structures calculated numerically by solving Eq. (\ref{Eq:AppTDGL}) supplemented with a Maxwell equation 
 are shown in Fig. \ref{f2} for different values of magnetic induction. In these plots and below we normalize radial coordinate $R=r\lambda$ by the  
 London penetration length $\lambda=\phi_0/\sqrt{-32\pi^3 K\alpha/\beta}$. 
 
  The stationary motion of vortices with a constant velocity $\bm U$ 
  is determined by the balance between Lorentz force on the vortex line 
 $\bm f_\nn{ext}=\bm j_\nn{tr} \times\int\bm Bd^2{\bm R}/c$ and a viscous friction  
  $\bm f_\nn{env}=-\eta\bm U$, where $\eta$ is the vortex viscosity. 
  To calculate $\eta$ we use the TDGL theory assuming that the order parameter and magnetic field 
   can be approximated by their equilibrium distributions transformed to the moving coordinate frame according to the 
   Galilean transformation of the fields. To find the electric field generated by moving vortex array we employ again the circular cell approximation.
  By assuming that ${\bm U}\parallel {\bm x}$, we take scalar potential in the form 
$\varphi({\bm R})=\phi_0 U\sin\theta b(r)/(2\pi c\lambda)$. 
 Taking into account above mentioned equation describing screening of electric field, we arrive to the non-homogeneous linear equation for the 
  scalar potential where the source term is generated by the moving vortex phase singularity 
 \begin{align}\label{varphi}
 b^{\prime\prime}+\frac{b^\prime}{r}-\frac{b}{r^2}=u\kappa^2f^2\Big(b+\frac{1}{r}\Big),
 \end{align}
where $\kappa=\lambda/\xi$ is a GL parameter.
Equation is supplemented by the boundary conditions at $r=0$ and $r=r_\nn{c}$.
 The former results from the regularity criterion $b(0)=0$. The latter can be obtained from the condition that the 
 average electric field $\mathcal{E}$ should satisfy a general relation  
 $c\mathcal{E}=\mathcal{B}\times\bm U$. Using  this constraint one gets boundary condition $b(r_\nn{c})=-1/r_\nn{c}$.  
 Eq. (\ref{varphi}) defines a non-equilibrium electric response of a moving vortex. The problem is parametrized by only one parameter, 
 $u$, which can change in wide interval controlling the electric field relaxation length 
  relative to the vortex core size as illustrated in Fig. \ref{f2} where the distribution of $b$ is shown for different magnetic fields and various values of $u$.   

To discuss resistive state in multiband superconductor, we first analyze forces driving the motion of a single vortex. Following the general procedure outlined in \cite{kopnin}, we obtain a general expression for the vortex viscosity within  a circular cell approximation
\begin{align}\label{fenv}
&\eta=2\pi\Gamma\psi_0^2\left[I_\nn{T}+I_\nn{BS}-\left(\frac{b^\prime r+b}{u\kappa^2r}\right)_{r_\nn{c}}+\int\limits_0^{r_\nn{c}}\frac{a^{\prime2}d r}{u\kappa^2r}\right],
\end{align}
where $I_\nn{T}=\int_0^{r_\nn{c}} f^{\prime2}r d r$ and 
$I_\nn{BS}=\int_0^{r_\nn{c}} f^2(b+1/r)d r$.
The last term in Eq. (\ref{fenv}) contains
the dimensionless magnetic flux $a(r)=\kappa^2\int_0^rdr_1r_1B(r_1)/H_{\nn{c}2}$. The flux-flow resistivity is given by the standard relation 
 \begin{equation}
\rho_f= \phi_0{\cal B} /(c^2\eta).
\end{equation}   
  
 For dilute vortex lattices ${\cal B}\ll H_{\nn{c}2}$ and $\kappa\gg1$, the last two terms in Eq. (\ref{fenv}) can be neglected. 
Therefore, for isolated vortices in extreme type-2 superconductors, the first two terms dominate. 
 One of them $I_\nn{T}$ is the Tinkham's part characterizing the order parameter relaxation and the other $I_\nn{BS}$
  takes into account Ohmic losses generated by normal currents flowing around moving vortices as discussed above. 
The former is fixed by an equlibrium vortex structure so that $I_\nn{T}\approx0.28$ while 
the value of $I_\nn{BS}$ is sensitive to parameter $u$.

In multiband superconductors where $u$ can be made arbitrary small, $I_\nn{BS}$ can be strongly enhanced
 compared to the single-component case where $I_\nn{BS}\approx 0.23$ \cite{kopninivlevKalatsky}.  
As a result, the contribution of Ohmic losses to the overall vortex energy dissipation is enhanced which means that 
the environment becomes more viscous for moving composite vortices as 
compared to the single-component case. 

 \begin{figure}[t!]
 \includegraphics[trim=5 250 155 420,clip,width=1.0\linewidth]{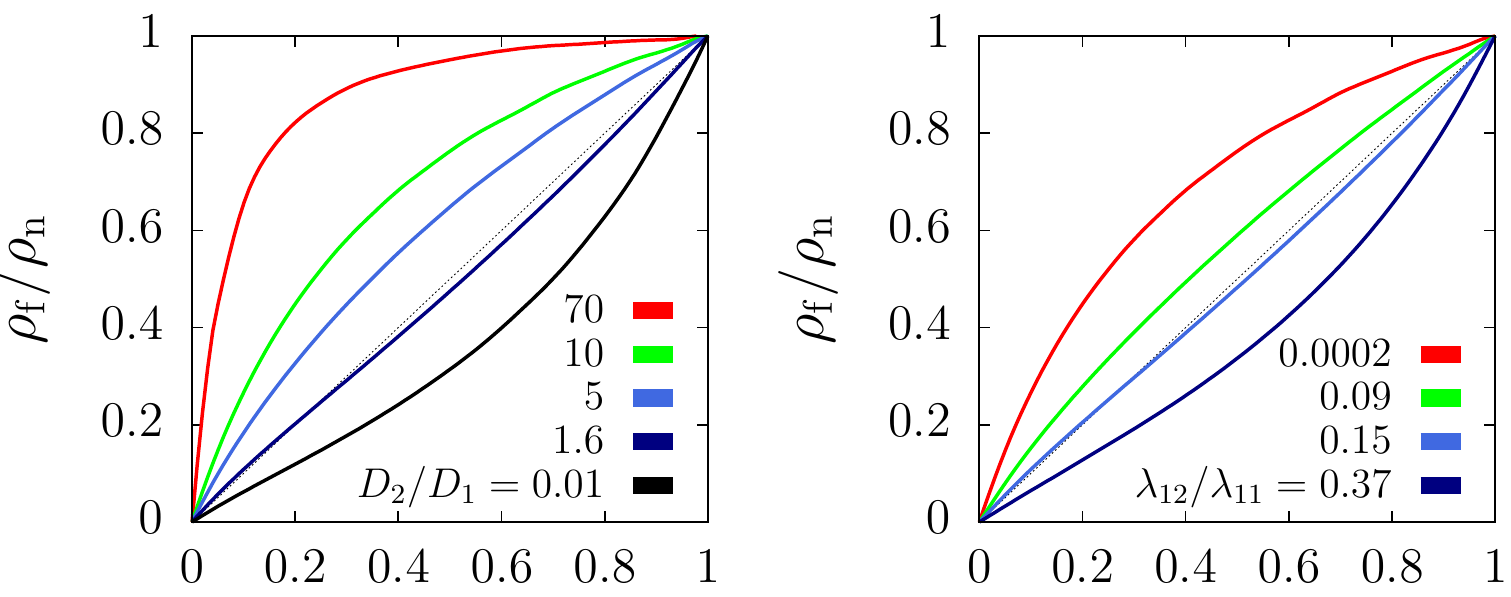} 
 \caption{\label{f3} 
 (Color online) Flux-flow resistivity \textit{vs} magnetic field. Left: As diffusion coefficient in the passive weaker band, 
 $D_2$, increases ($u=5.7,\ 2.7,\ 1.3,\ 0.7,\ 0.1$, correspondingly). Right: As interband interaction constant 
 $\lambda_{12}$ increases ($u=0.7,\ 1.7,\ 2.6,\ 4.9$,  correspondingly). Thin dotted line is BS result. 
}
 \end{figure}

 For weak fields ${\cal B}\ll H_{\nn{c}2}$ and large $\kappa\gg 1$, one gets
\begin{align}\label{rhoff1}
&\frac{\rho_\nn{f}}{\rho_\nn{n}}=\frac{2}{u(I_\nn{T}+I_\nn{BS})}\frac{\mathcal{B}}{H_{\nn{c}2}}.
\end{align}
 The presence of parameter $u$ in the denominator signals a possibility to enhance magneto-resistance $\gamma$ due to the 
 strong delocalization of
 electric field generated by composite vortices when $u\to 0$. This conclusion is confirmed by numerically calculated
  flux-flow resistivity curves. As shown in Fig. \ref{f3}, 
 the dependencies $\rho_f({\cal B})$ can have much steeper slopes at ${\cal B}\ll H_{c2}$ as compared to
  the usual BS law shown by the dotted line. 

 Within the circular cell approximation, Eq. (\ref{rhoff1}) 
 can be applied to calculate the flux-flow resistivity at arbitrary fields $0<\mathcal{B}<H_{\nn{c}2}$.
 At that, denominator in Eq. (\ref{rhoff1}) should be supplemented by the last two terms in Eq. (\ref{fenv}).
 Such an approach is consistent with well-known analytical asymptote near $H_{\nn{c}2}$
 \begin{align}\label{rhoff2}
\frac{\rho_\nn{f}}{\rho_\nn{n}}=1-\frac{u\kappa^2}{\beta_\nn{A}(2\kappa^2-1)+1}\left(1-\frac{\mathcal{B}}{H_{\nn{c}2}}\right),
\end{align}
where $\beta_\nn{A}=1.16$ is Abrikosov parameter \cite{kleiner}. In single-band large-$\kappa$ superconductors, 
the slope of the dependence (\ref{rhoff2}) is $(H_{c2}/\rho_\nn{n}) \partial_{\cal B}\rho_\nn{f} = 2.5$ meaning that the curve $\rho_f({\cal B})/\rho_n$ goes below the 
BS line. In the two-band case shown in Fig. \ref{f3}, we obtain much more diverse behaviour in qualitative agreement with recent experimental data \cite{exp3,exp4} with enhanced magneto-resistance.
  
 Energy dissipation by moving vortices is key limiting factor for practical applications of superconductivity. 
 The diverse flux-flow behaviour of multiband superconductors has not been 
  properly described by the existing theories developed for conventional single-band superconductors. 
 Qualitatively the reason for disagreement was that the size of a non-equilibrium domain  
 with localized electric field has been assumed to be proportional to the vortex core size, 
 as in the pioneering work by Bardeen and Stephen. 
 As we have demonstrated above this assumption is qualitatively incorrect for complex composite vortices
 in multiband superconductors formed by coexisting condensates in different superconducting bands. 
 For these objects, an interplay of microscopic parameters such as diffusion coefficients and pairing constants in
  different bands can lead to the large variations in electric field relaxation length relative to the vortex core size.
 As a result, electric field can be stretched strongly outside the vortex cores dramatically enhancing the Ohmic losses and the overall energy dissipation by moving fluxes 
\footnote{Another mechanism which can affect flux-flow resistivity in  multiband superconductors is splitting of moving composite vortices into a non-cocentered fractional ones under the action of an external current. This effect however should be diminished by
interband Josephson coupling since fractional vortices
attract each other linearly at the length scales larger than Josephson length (see e.g. \cite{babaev2}). In particularly
it should not be important  in the region of our interest: near $T_\nn{c}$
because the Josephson length does not diverge when $T\to T_\nn{c}$.}.

The unusual flux-flow phenomena that we discuss here is one of the possible examples of resistive 
states in superconductors. 
Broad range of non-stationary phenomena which should be strongly affected by multiband effects include the 
formation of resistive states in narrow superconducting channels such as the mesoscopic wires of 
the width compared to the coherence length \cite{WattsTobin,Ivlev}.
Such systems are quite important for technological applications in superconducting photon detectors \cite{Detectors}.   
Near critical temperature they have rather complicated behaviour strongly affected 
by the value of parameter $u$ in TDGL theory. 
Different non-stationary regimes 
including the growth of critical superconducting nucleus and the formation of phase slip centres
in homogeneous channels \cite{WattsTobin} and near the boundaries \cite{Ivlev}  
have been thoroughly investigated for single-band compounds with the fixed value of $u=5.79$. 
Generically different regimes with much smaller 
values of $u$ become accessible in multiband superconductors reported in the present paper. 
This possibility opens potentially interesting 
directions of studying non-equilibrium current-carrying states in thin films and wires of multiband superconductors. 

\begin{acknowledgements}
The work was supported by
the Swedish Institute, Estonian Ministry of Education and Research (grant PUTJD141), Goran Gustafsson Foundation  and by the Swedish
Research Council grant 642-2013-7837.
\end{acknowledgements}
\bibliography{viscosity2}
\end{document}